\documentclass[11pt]{article}
\usepackage[margin=1in]{geometry}
\usepackage{amsmath,amssymb,amsfonts}
\usepackage{graphicx}
\usepackage{tikz-cd}
\usepackage{multicol}
\usepackage{authblk}
\usepackage{url}
\usepackage{bm}
\usepackage[numbers,sort&compress]{natbib}
\title{Variational Quantum Eigensolver-Based Quantum Bootstrap Embedding for Molecules}

\author[1]{Derek Peng}

\affil[1]{\small The Pingry School, Basking Ridge, NJ, USA}

\date{}

\begin{document}

\maketitle

\begin{abstract}
Simulating strongly correlated molecular systems on near-term quantum hardware remains challenging because current hardware offers limited quantum volume and moderate-fidelity qubits. One way to address this challenge is through bootstrap embedding (BE). Bootstrap embedding partitions molecules into smaller fragments that are then embedded in the ``bath'' of other fragments iteratively. Bootstrap embedding is appealing for quantum simulation because fragmentation reduces the qubit requirements of each fragment problem. In this work, we develop a quantum bootstrap embedding (QBE) workflow that uses variational quantum eigensolver (VQE) fragment solvers and examines the algorithmic choices that determine the success of the overall VQE-QBE procedure. To improve efficiency, we introduce FastAdaptVQE, a sparse-matrix-accelerated form of the adaptive variational quantum eigensolver (ADAPT-VQE) that replaces symbolic commutator evaluation with direct statevector calculations, and MatrixFreeAdaptVQE, a matrix-free extension that removes the sparse-matrix memory bottleneck that arises for larger fragments. We also explore modifying the ADAPT-VQE operator-selection step by replacing the purely greedy choice with a lookahead strategy. Benchmarks on H$_4$, F$_2$, and selected geometries of H$_6$ reach chemical accuracy, within 1 kcal/mol of BE results obtained with a full configuration interaction (FCI) solver. For $n$-butane, density-matching residuals converged, but energy errors remained outside chemical accuracy. These results show that combining QBE with VQE can accurately compute molecular energies, while density convergence alone does not ensure energy accuracy. This work lays the foundation for extending these energy calculations to larger molecular systems and quantum materials on near-term quantum hardware.
\end{abstract}


\section{Introduction}

Quantum computing offers a compelling route to \emph{ab initio} electronic structure because molecular wave functions live in Hilbert spaces whose dimensions grow combinatorially with active-space size \cite{AspuruGuzik2005,McArdle2020,Bauer2020}. Present devices, however, remain in the noisy intermediate-scale quantum (NISQ) regime, in which noise, finite coherence, limited connectivity, and finite sampling restrict the circuit width (number of qubits) and depth that can be executed reliably \cite{Preskill2018,Bharti2022,Cerezo2021,Endo2021}. A useful hardware-level summary of this constraint is quantum volume, which measures the largest random circuit of a given width and depth that a device can implement successfully \cite{Cross2019}. For molecular simulation, useful near-term algorithms therefore must reduce qubit counts, circuit depth, and measurement requirements while preserving chemically meaningful electronic structure information \cite{Cao2019,Tilly2022}.

Hybrid quantum--classical approaches are especially attractive under these constraints. The variational quantum eigensolver (VQE) replaces long coherent time evolution by iterative preparation and measurement of a parameterized quantum state, making it a leading NISQ strategy for molecular simulation \cite{Peruzzo2014,McClean2016,Kandala2017,OMalley2016,Tilly2022}. VQE performance depends strongly on the ansatz, optimizer, measurement strategy, and noise-mitigation protocol \cite{McClean2016,Cerezo2021,Endo2021,Tilly2022}. Chemically motivated ansatze such as unitary coupled cluster singles and doubles (UCCSD) connect VQE to classical wave-function theory, while adaptive variants such as the adaptive variational quantum eigensolver (ADAPT-VQE) build compact, system-specific ansatze by selecting operators from a pool according to energy gradients \cite{Taube2006,Romero2018,Grimsley2019,Claudino2020}. Subsequent ADAPT variants have explored qubit-operator and qubit-excitation operator pools, symmetry-aware pools, batched growth, and circuit-depth reductions, reflecting the central role of operator selection in near-term quantum chemistry \cite{Tang2021QubitAdapt,Yordanov2021,Shkolnikov2023,Sapova2022Batched,Anastasiou2024Tetris}. Even with these improvements, direct VQE calculations become increasingly difficult as the number of orbitals and electrons increases because larger active spaces require more qubits, larger operator pools, and more demanding optimization and measurement resources \cite{Cao2019,McArdle2020,Bauer2020}. These limitations motivate algorithmic strategies that shrink the quantum problem itself.

Embedding methods provide one of the most direct ways to address this size barrier. Rather than solving the full many-electron problem in one calculation, embedding partitions a large system into smaller fragments and represents the surrounding environment through a bath, mean-field, or self-consistent matching construction \cite{Knizia2013,Wouters2016,Sun2016QuantumEmbedding}. Density matrix embedding theory and related quantum embedding methods have shown that accurate local correlated calculations can be embedded into a lower-cost description of the full system, and they have also motivated hybrid quantum--classical embedding workflows in which quantum processors solve only the embedding subproblems \cite{Rubin2016,Mineh2022,Li2022}. Within this broader family, bootstrap embedding (BE) was introduced as a self-consistent overlapping-fragment framework in which matching conditions between neighboring fragments reduce boundary errors that limit simpler fragmentation methods \cite{Welborn2016}.

BE was subsequently extended from model Hamiltonians to molecular systems, where its accuracy was shown to improve rapidly with fragment size and where overlapping fragments naturally generated matching conditions between neighboring subsystems \cite{Ye2019,Ye2019Atom}. Later developments demonstrated that BE can be implemented with favorable low scaling for large molecules and can recover the vast majority of the correlation energy of high-level classical methods such as coupled cluster singles and doubles (CCSD) for sizeable systems and extended basis sets \cite{Ye2020,Tran2024}. Related periodic formulations and software developments have further extended the scope of the approach beyond finite molecules \cite{Meitei2023Periodic,Cho2025QuEmb}. These results establish BE as a systematically improvable classical embedding strategy.

The same fragmentation logic makes BE conceptually attractive for quantum computing. If a large molecular Hamiltonian can be decomposed into overlapping embedded fragments, each fragment Hamiltonian can be assigned to a smaller quantum register than would be required for the full system. Liu \textit{et al.} made this connection explicit in their formulation of quantum bootstrap embedding (QBE), which uses embedding to tailor fragment sizes to the limited size of current quantum machines while still targeting the electronic structure of the full molecule \cite{Liu2023QBE}. Recent sample-based QBE work has further reinforced the relevance of QBE to hardware-facing electronic-structure workflows by combining QBE with sample-based diagonalization on a real superconducting processor \cite{Bierman2026}.

For QBE to become practical on near-term devices, the fragment solver must itself be compatible with noisy hardware. This requirement makes shallow, flexible variational solvers natural candidates for QBE fragments \cite{Peruzzo2014,McClean2016,Cerezo2021,Tilly2022}. The original QBE formulation focused primarily on the embedding formalism and on coherent matching strategies based on SWAP tests and amplitude amplification; although approximate VQE calculations were discussed, a detailed treatment of BE combined with VQE was left outside of that study \cite{Liu2023QBE}. Thus, a systematic study of how to make a molecular QBE pipeline work with VQE fragment solvers remains needed \cite{Liu2023QBE,Bierman2026}.

In this paper, we develop a self-consistent QBE workflow for molecular energy calculations on near-term quantum hardware by combining bootstrap embedding with VQE fragment solvers. Since the quality of the coupled VQE--QBE approach depends not only on the fragmentation itself but also on the efficiency and accuracy of the inner variational solver, we study the algorithmic choices that govern its overall performance. Specifically, we use adaptive fragment solvers based on ADAPT-VQE \cite{Grimsley2019}; introduce FastAdaptVQE, a sparse-matrix accelerated implementation of ADAPT-VQE that replaces symbolic commutator evaluation with direct statevector linear algebra; introduce MatrixFreeAdaptVQE, a matrix-free extension that applies operators directly to the statevector rather than storing large sparse matrices; and incorporate a look-ahead operator-selection strategy that improves ansatz construction by considering downstream optimizations rather than making purely greedy local choices. In this progression, FastAdaptVQE removes the symbolic gradient-construction bottleneck, whereas MatrixFreeAdaptVQE removes the sparse-matrix memory bottleneck that emerges at larger fragment sizes. Using noiseless classical statevector simulations, we benchmark the method on H$_4$, H$_6$, F$_2$, and $n$-butane. For H$_4$, F$_2$, and selected H$_6$ geometries, the density-matching error decreases over successive bootstrap embedding iterations and the energy converges to within chemical accuracy relative to bootstrap embedding with FCI; for $n$-butane, the density-matching residuals converge, but the energy errors remain outside chemical accuracy. To our knowledge, this work is the first to apply a bootstrap embedding framework with VQE fragment solvers for molecular energy calculations, complementing earlier QBE and sample-based QBE studies \cite{Liu2023QBE,Bierman2026}.

\section{Methods}

\subsection{Overview of the VQE--QBE workflow}

\begin{figure}[htbp]
    \centering
    \includegraphics[width=1.0\textwidth]{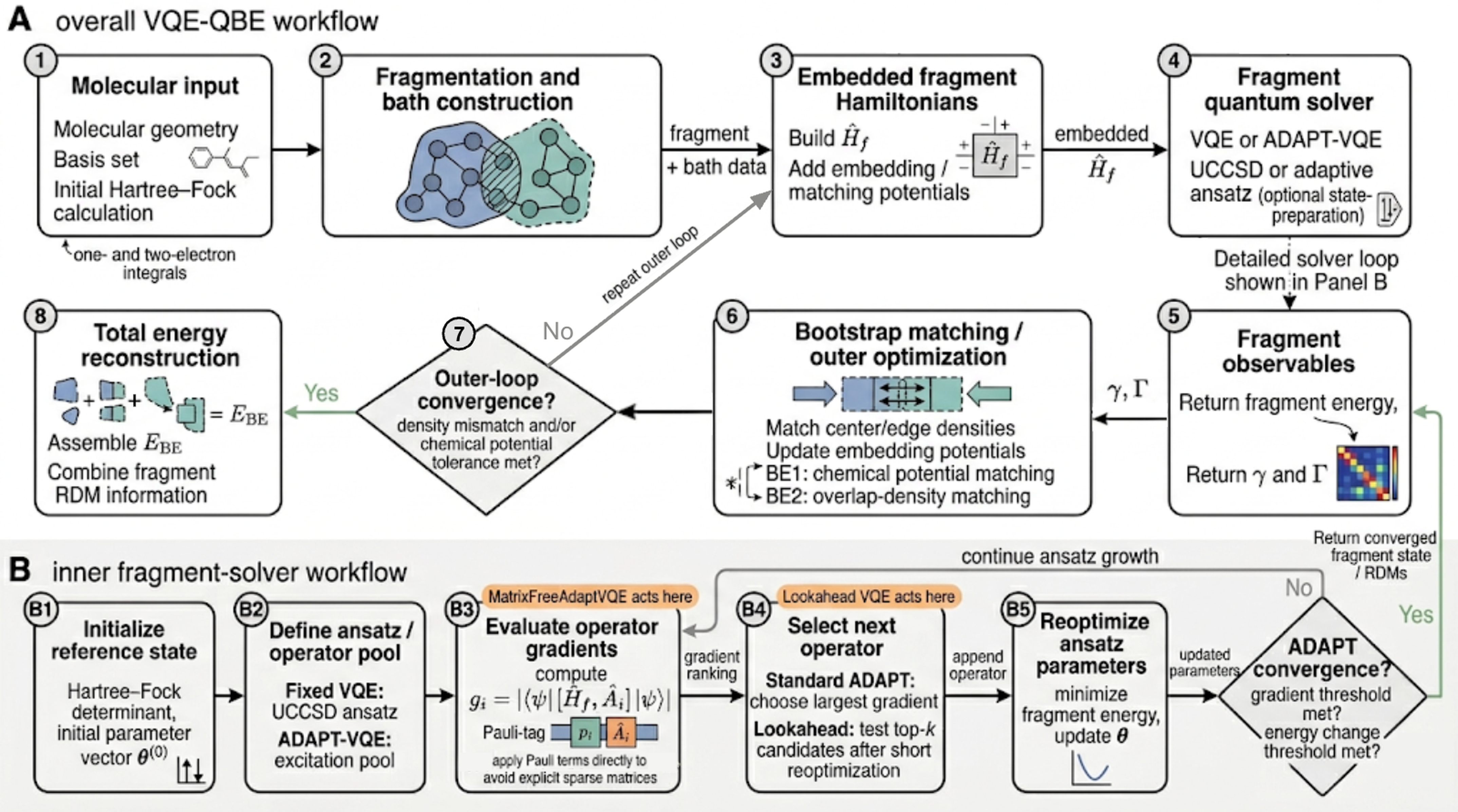}
    \caption{
Workflow of the variational quantum eigensolver--based quantum bootstrap embedding (VQE--QBE) procedure. \textbf{A} The outer bootstrap-embedding cycle partitions the molecule into fragments, constructs embedded fragment Hamiltonians, solves each fragment, returns fragment reduced density matrices, updates the matching potentials, and iterates to self-consistency before reconstructing the total embedded energy. \textbf{B} Inner fragment-solver loop. Each embedded fragment is solved with VQE or ADAPT-VQE. In the adaptive solver, candidate gradients are evaluated, one operator is appended, and all variational parameters are reoptimized until convergence. FastAdaptVQE and MatrixFreeAdaptVQE change the statevector representation used for candidate screening, while lookahead changes the operator-selection procedure.
}
    \label{fig:pipeline}
\end{figure}

The overall calculation is a nested quantum--classical optimization in which bootstrap embedding (BE) defines a sequence of embedded fragment Hamiltonians and a variational quantum eigensolver (VQE) is used as the fragment solver
\cite{Peruzzo2014,McClean2016,Welborn2016,Liu2023QBE}. For each fragment \(A\),
BE \cite{Welborn2016,Ye2019,Ye2019Atom,Ye2020,Tran2024,Liu2023QBE,Bierman2026} constructs an embedded Hamiltonian \(\hat{H}^{\mathrm{emb}}_A(\boldsymbol{\lambda},\mu)\), where \(\boldsymbol{\lambda}\) denotes the matching potentials and \(\mu\) is the global chemical potential. Each embedded fragment problem is then solved with a VQE to obtain the fragment energy and reduced density matrices (RDMs), which are returned to the embedding layer to update \(\boldsymbol{\lambda}\) and \(\mu\). This outer BE loop is iterated until the matching loss and total energy have converged. A schematic of the overall VQE--QBE workflow is shown in Figure \ref{fig:pipeline}.

This structure separates the method into an embedding layer (BE self-consistency) and a fragment-solver layer (VQE/ADAPT-VQE and the variants introduced here). Throughout this work, the fragment problems are solved via classical simulations of the VQE ansatz on classical hardware.

\subsection{Bootstrap Embedding}

The outer self-consistency problem is solved with bootstrap embedding (BE)
\cite{Welborn2016,Ye2019,Ye2019Atom,Ye2020,Tran2024,Liu2023QBE}. BE partitions
the molecule into overlapping fragments and enforces consistency conditions between neighboring fragments through an outer optimization over local one-body embedding potentials. The starting point is a full-molecule restricted Hartree--Fock calculation. For the H$_4$ and F$_2$ examples studied here, the default local-orbital basis is the symmetrically orthogonalized L\"owdin basis $ W = S^{-1/2} $ \cite{Lowdin1950}, where $S$ is the atomic-orbital (AO) overlap matrix. If $C_{\mathrm{occ}}$ denotes the occupied molecular orbitals, the occupied local orbitals are $C_{\mathrm{loc}} = W^{\mathsf T} S C_{\mathrm{occ}}, $ and the occupied-space density matrix in the local basis is $ D = C_{\mathrm{loc}} C_{\mathrm{loc}}^{\mathsf T}. $ For fragment $A$ with local-orbital index set $F_A$ and environment $E_A$, the bath is obtained from a Schmidt decomposition of the environment block $ D_{E_AE_A} = U_A \Lambda_A U_A^{\mathsf T}, $ where the first $E_A$ represents the row index set,  while the second $E_A$ represents the column index set. Bath orbitals are the eigenvectors whose Schmidt eigenvalues satisfy
\begin{equation}
\tau_{\mathrm{bath}} < |\lambda_\beta^{(A)}| < 1-\tau_{\mathrm{bath}},
\end{equation}
with $\tau_{\mathrm{bath}} = 10^{-10}$ in the H$_4$ and F$_2$ benchmark controls. The embedded fragment space is spanned by the fragment orbitals together with the retained bath orbitals,
$
T_A = [I_{F_A}\ \ U_A^{\mathrm{bath}}],
$
where $I_{F_A}$ denotes the fragment-orbital identity block and $n_{\mathrm{bath}}^{(A)}$ is the number of retained bath orbitals, so each embedded fragment contains $|F_A| + n_{\mathrm{bath}}^{(A)}$ spatial orbitals.

The atom-based fragmentation used here follows the molecular BE construction in which atom groups are connected by local geometry. Candidate neighboring groups are searched within 3.5~\AA, with bond cutoffs of 1.8~\AA\ for general atoms and 1.2~\AA\ for hydrogen. Each fragment contains an origin motif, center orbitals, and edge orbitals. The center orbitals are the orbitals whose RDM elements contribute to the energy reconstruction for that fragment; the edge orbitals are matched to the corresponding center orbitals of a neighboring reference fragment. For fragment $A$, we write the embedded Hamiltonian as
\begin{equation}
\hat{H}^{\mathrm{emb}}_A(\boldsymbol{\lambda},\mu)
=
\hat{H}^{(0)}_A + \hat{V}^{\mathrm{BE}}_A(\boldsymbol{\lambda},\mu),
\end{equation}
where $\hat{H}^{(0)}_A$ is the bare fragment Hamiltonian and $\hat{V}^{\mathrm{BE}}_A$ is the BE one-body potential determined by the current matching multipliers $\boldsymbol{\lambda}$ and global chemical potential $\mu$. It is convenient to write this potential as
\begin{equation}
\hat{V}^{\mathrm{BE}}_A(\boldsymbol{\lambda},\mu)
=
\sum_{xy} v^{(A)}_{xy}(\boldsymbol{\lambda},\mu)\, \hat{a}_x^{\dagger} \hat{a}_y,
\end{equation}
where the Hermitian matrix $v^{(A)}(\boldsymbol{\lambda},\mu)$ collects the fragment-local matching terms and the chemical-potential contribution used to enforce the correct total electron count. Each fragment contains a \emph{center} region, whose observables contribute to the molecular reconstruction, and an \emph{edge} region, which overlaps with neighboring fragment centers and supplies the matching conditions.

For an overlapping pair of fragments $A$ and $B$, let $\mathcal{O}_{AB}$ denote the shared orbital region represented as the edge of one fragment and the center of the other. The BE2 matching condition is implemented as the root-mean-square (RMS) norm of an explicit error vector that contains all unique edge-minus-center one-particle density elements together with the chemical-potential residual. Here $\gamma^A$ denotes the spin-summed spatial-orbital one-particle RDM returned by fragment $A$:
\begin{equation}
\mathbf{r}_{\mathrm{BE2}}
=
\Big(
\{\gamma_{pq}^{A,\mathrm{edge}}-\gamma_{pq}^{B,\mathrm{center}}\}_{A<B,\ p\le q \in \mathcal{O}_{AB}},
\Delta N
\Big),
\end{equation}
with
\begin{equation}
\Delta N
=
\sum_{A=1}^{N_{\mathrm{frag}}}\sum_{p \in C_A} \gamma_{pp}^{A} - N_e
\end{equation}
where $N_e$ is the target total electron count, and
\begin{equation}
\mathcal{L}_{\mathrm{BE2}}
=
\sqrt{\frac{1}{n_r}\,\mathbf{r}_{\mathrm{BE2}}^{\mathsf T}\mathbf{r}_{\mathrm{BE2}}},
\end{equation}
where $C_A$ denotes the center orbitals of fragment $A$ and $n_r$ is the length of the error vector.

For BE1 calculations, where fragments do not share a spatial overlap, the matching reduces to the scalar chemical-potential residual
\begin{equation}
\mathbf{r}_{\mathrm{BE1}}(\mu)
=
\sum_{A=1}^{N_{\mathrm{frag}}}\sum_{p \in C_A}\gamma_{pp}^{A} - N_e,
\qquad
\mathcal{L}_{\mathrm{BE1}} = |\mathbf{r}_{\mathrm{BE1}}|.
\end{equation}
Thus, the role of the fragment solver is not only to minimize an isolated fragment energy, but also to return RDMs accurate enough to make the outer matching conditions stationary.

After convergence of the outer loop, the total molecular energy is reconstructed from the fragment RDMs using the cumulant-based BE energy expression employed in classical molecular BE literature \cite{Ye2019,Ye2020,Tran2024},
\begin{equation}
E_{\mathrm{BE}}
=
E_{\mathrm{HF}}
+
\sum_{pq} F_{pq}\,\Delta\gamma_{qp}
+
\frac{1}{2}\sum_{pqrs} (pq|rs)\,K^{\mathrm{approx}}_{pqrs},
\end{equation}
where $E_{\mathrm{HF}}$ is the Hartree--Fock energy, $F_{pq}$ are the spin-summed Fock-matrix elements in the spatial-orbital basis, $(pq|rs)$ are the two-electron repulsion integrals in chemists' notation, $\Delta\gamma = \gamma - \gamma^{\mathrm{HF}}$ is the correlation correction to the one-particle density matrix. In the spin-summed closed-shell convention used here, the fragment cumulant is obtained from the fragment two-particle RDM as
\begin{equation}
K^{(A)}_{pqrs}
=
\Gamma^{(A)}_{pqrs}
-\left(
\gamma^{(A)}_{pq}\gamma^{(A)}_{rs}
-\frac{1}{2}\gamma^{(A)}_{ps}\gamma^{(A)}_{rq}
\right).
\end{equation}
The approximate molecular cumulant $K^{\mathrm{approx}}$ is assembled by projecting the center contribution of each $K^{(A)}$ into the molecular basis and summing the center-weighted terms, following the molecular BE reconstruction \cite{Ye2019,Ye2020,Tran2024}. The same fragment partition, local basis, bath threshold, and BE energy expression are used for all fragment solvers compared in this study so that differences in the final embedded energy can be attributed to the fragment solver rather than to a changed outer embedding definition.

The outer optimization variables are the local matching potentials together with the global chemical potential,
\begin{equation}
\mathbf{x} = (\boldsymbol{\lambda},\mu),
\qquad
\mathbf{x}^{(0)}=\mathbf{0}.
\end{equation}
The benchmark path uses the default quasi-Newton outer solver with a Broyden rank-one update \cite{Broyden1965}. At iteration $t$, one forms the BE error vector $\mathbf{r}^{(t)}=\mathbf{r}(\mathbf{x}^{(t)})$ and solves
\begin{equation}
B^{(t)} \Delta \mathbf{x}^{(t)} = -\mathbf{r}^{(t)},
\qquad
\mathbf{x}^{(t+1)} = \mathbf{x}^{(t)} + \alpha_t \Delta \mathbf{x}^{(t)},
\end{equation}
where $B^{(t)}$ is a Jacobian approximation and $\alpha_t$ is a line-search parameter. The initial Jacobian is taken from the Hartree--Fock response and is updated by a rank-one Broyden step,
\begin{equation}
B^{(t+1)}
=
B^{(t)}
+
\frac{
\left(
\Delta\mathbf{r}^{(t)} - B^{(t)}\Delta\mathbf{x}^{(t)}
\right)
\Delta\mathbf{x}^{(t)\mathsf T}
}{
\Delta\mathbf{x}^{(t)\mathsf T}\Delta\mathbf{x}^{(t)}
}.
\end{equation}
In the benchmark controls, the outer BE tolerance is $10^{-6}$ and the maximum number of BE iterations is 50.

\subsection{Variational Quantum Eigensolver}

Each embedded fragment is solved with VQE \cite{Peruzzo2014,McClean2016,Cao2019,Tilly2022}. For fragment $A$, we write the embedded electronic Hamiltonian in an orthonormal \emph{spin-orbital} basis as
\begin{equation}
\hat{H}^{\mathrm{emb}}_A
=
E^{(A)}_0
+
\sum_{ij} h^{(A)}_{ij}\, \hat{a}_i^\dagger \hat{a}_j
+
\frac{1}{2}\sum_{ijkl} g^{(A)}_{ijkl}\,
\hat{a}_i^\dagger \hat{a}_j^\dagger \hat{a}_l \hat{a}_k,
\label{eq:embedded-hamiltonian}
\end{equation}
where $E^{(A)}_0$ is the fragment-specific scalar offset; $h^{(A)}_{ij}$ and $g^{(A)}_{ijkl}$ are the one- and two-electron integrals of the embedded fragment Hamiltonian; $i,j,k,l$ label spin orbitals; and $\hat{a}_i^\dagger$ and $\hat{a}_i$ are the corresponding fermionic creation and annihilation operators. In practice, $\hat{H}^{\mathrm{emb}}_A$ contains both the bare fragment Hamiltonian and the current BE one-body embedding potential. The same embedded Hamiltonian is used for fragment optimization and for the post-optimization evaluation of fragment RDMs returned to the embedding layer.

The fermionic Hamiltonian is mapped to qubits with the Jordan--Wigner transformation, which results in the Pauli decomposition
\begin{equation}
\hat{H}^{\mathrm{emb}}_A = \sum_{\ell} c^{(A)}_{\ell} \hat{P}_{\ell},
\end{equation}
where $c^{(A)}_{\ell}$ are the scalar coefficients of the Pauli terms and each $\hat{P}_{\ell}$ is a tensor product of Pauli operators. The VQE ansatz is written as
\begin{equation}
\begin{aligned}
|\psi_A(\boldsymbol{\theta})\rangle
&= \hat{U}_A(\boldsymbol{\theta}) |\phi_A^{\mathrm{ref}}\rangle, \\
\hat{U}_A(\boldsymbol{\theta})
&= e^{\theta_{N_p} \hat{\kappa}_{N_p}} \cdots
e^{\theta_2 \hat{\kappa}_2} e^{\theta_1 \hat{\kappa}_1},
\end{aligned}
\end{equation}
where $|\phi_A^{\mathrm{ref}}\rangle$ is the fragment reference determinant, $\{\hat{\kappa}_j\}$ are anti-Hermitian generators, and $\boldsymbol{\theta}=(\theta_1,\dots,\theta_{N_p})$ is the variational parameter vector with $N_p$ variational amplitudes. The fragment energy is given by the expectation value
\begin{equation}
E_A(\boldsymbol{\theta})
=
\langle \psi_A(\boldsymbol{\theta}) | \hat{H}^{\mathrm{emb}}_A | \psi_A(\boldsymbol{\theta}) \rangle,
\end{equation}
assuming the wave functions are normalized. The classical optimizer updates $\boldsymbol{\theta}$ to minimize $E_A(\boldsymbol{\theta})$.

All ansatzes used here conserve particle number and $S_z$, the z-component of the total electronic spin angular momentum. Each fragment solution therefore remains in the symmetry sector of its reference determinant. Unless otherwise noted, the fragment reference is taken from the Hartree--Fock solution of the current embedded fragment Hamiltonian in the fragment+bath basis, so that the VQE initialization is consistent with the effective Hamiltonian used in the current outer BE iteration.

For the specific VQE ansatze, we use a unitary coupled-cluster singles and doubles (UCCSD) state built on the fragment Hartree--Fock reference,
\begin{equation}
|\psi_A^{\mathrm{UCCSD}}(\boldsymbol{\theta})\rangle
=
\exp\!\left[\hat{T}(\boldsymbol{\theta})-\hat{T}^{\dagger}(\boldsymbol{\theta})\right]
|\phi_A^{\mathrm{HF}}\rangle,
\end{equation}
with
\begin{equation}
\begin{aligned}
\hat{T}(\boldsymbol{\theta}) &= \hat{T}_1 + \hat{T}_2, \\
\hat{T}_1 &= \sum_{ia} t_i^a \hat{a}_a^\dagger \hat{a}_i, \\
\hat{T}_2 &= \frac{1}{4}\sum_{ijab} t_{ij}^{ab}
\hat{a}_a^\dagger \hat{a}_b^\dagger \hat{a}_j \hat{a}_i.
\end{aligned}
\end{equation}
Here $i,j$ denote occupied spin orbitals of the reference determinant, $a,b$ denote virtual spin orbitals, and $\boldsymbol{\theta}=\{t_i^a,t_{ij}^{ab}\}$ are the variational amplitudes. The UCCSD form provides a chemically motivated reference parameterization, whereas the adaptive ansatzes described below select only a subset of such excitations.

At the circuit level, the fixed-ansatz controls use the Hartree--Fock reference in the Jordan--Wigner qubit basis together with the standard particle-number- and $S_z$-conserving singles-and-doubles unitary coupled-cluster (UCC) excitation manifold. The resulting circuit is transpiled into the gate basis $\{\mathrm{cx},\mathrm{rz},\mathrm{sx},x\}$ before the variational optimization.

The derivatives of the energy with respect to the variational parameters are used by the classical optimizer to determine the parameter updates. For a general differentiable ansatz,
\begin{equation}
\frac{\partial E_A}{\partial \theta_j}
=
2\,\mathrm{Re}\!
\left\langle
\frac{\partial \psi_A}{\partial \theta_j}
\middle|
\hat{H}^{\mathrm{emb}}_A
\middle|
\psi_A
\right\rangle.
\end{equation}
Conceptually, these derivatives enter a local parameter update of the form
\begin{equation}
\boldsymbol{\theta}^{(t+1)}
=
\boldsymbol{\theta}^{(t)} - \eta^{(t)} \nabla_{\boldsymbol{\theta}} E_A(\boldsymbol{\theta}^{(t)}),
\label{eq:schematic-gradient-update}
\end{equation}
where $\eta^{(t)}$ denotes an effective optimizer-dependent step size at iteration $t$. In other words, the energy gradient determines the direction in parameter space that lowers the fragment energy, and the optimizer uses that information to determine the parameters to be used in the next iteration. In the present implementation, Eq.~\eqref{eq:schematic-gradient-update} is only schematic: the actual update is generated by the sequential least-squares quadratic programming (SLSQP) optimizer \cite{Kraft1988SQP,Kraft1994TOMP} rather than by fixed-step gradient descent.

After variational convergence, the fragment solver returns the RDMs required by the embedding layer. To avoid ambiguity, we distinguish the spin-orbital Hamiltonian notation of Eq.~\eqref{eq:embedded-hamiltonian} from the \emph{spin-summed spatial-orbital} RDMs used for matching and energy reconstruction. We denote the spin-summed one-particle RDM by \(\gamma\) and the spin-summed two-particle RDM by \(\Gamma\):
\begin{equation}
\gamma_{pq}
=
\sum_{\sigma}
\langle \Psi_A | \hat{a}_{p\sigma}^{\dagger} \hat{a}_{q\sigma} | \Psi_A \rangle,
\end{equation}
\begin{equation}
\Gamma_{pqrs}
=
\sum_{\sigma\tau}
\langle \Psi_A | \hat{a}_{p\sigma}^{\dagger} \hat{a}_{r\tau}^{\dagger} \hat{a}_{s\tau} \hat{a}_{q\sigma} | \Psi_A \rangle,
\end{equation}
where $p,q,r,s$ now denote spatial orbitals and \(|\Psi_A\rangle\) is the optimized fragment state. These RDMs obey the contraction identity
\begin{equation}
\sum_{r} \Gamma_{pqrr} = (N_A-1)\gamma_{pq}
\end{equation}
for an $N_A$-electron fragment state.

\subsection{ADAPT-VQE}

To reduce ansatz size and circuit depth relative to a fixed UCCSD parameterization, we use ADAPT-VQE \cite{Grimsley2019,Claudino2020}. ADAPT constructs the wave function iteratively from an operator pool $\mathcal{P}$ rather than optimizing all single and double excitations simultaneously. After $m$ growth steps, the ansatz is
\begin{equation}
|\psi_A^{(m)}\rangle
=
e^{\theta_m \hat{A}_{\nu_m}} \cdots e^{\theta_2 \hat{A}_{\nu_2}} e^{\theta_1 \hat{A}_{\nu_1}}
|\phi_A^{\mathrm{ref}}\rangle,
\qquad
\hat{A}_{\nu_{\ell}} \in \mathcal{P},
\end{equation}
where the ordered list $(\hat{A}_{\nu_1},\dots,\hat{A}_{\nu_m})$ is selected adaptively. For the H$_4$ and F$_2$ calculations, the pool consists of non-generalized, spin-conserving one- and two-body excitation generators,
\begin{equation}
\hat{A}_i^a = \hat{a}_a^{\dagger} \hat{a}_i - \hat{a}_i^{\dagger} \hat{a}_a,
\qquad
\hat{A}_{ij}^{ab} = \hat{a}_a^{\dagger} \hat{a}_b^{\dagger} \hat{a}_j \hat{a}_i - \hat{a}_i^{\dagger} \hat{a}_j^{\dagger} \hat{a}_b \hat{a}_a.
\end{equation}
Because these generators are anti-Hermitian and preserve particle number and $S_z$, the ADAPT ansatz remains in the same symmetry sector as the reference determinant. The ansatz and derivative above use this anti-Hermitian source convention. The qubit implementation instead stores the corresponding Hermitian generator
\begin{equation}
\hat{G}_{\nu}=i\hat{A}_{\nu},
\qquad
e^{\theta\hat{A}_{\nu}}=e^{-i\theta\hat{G}_{\nu}}.
\end{equation}
Making this conversion explicit is important because the two conventions give different-looking, but equivalent, commutator formulas.

For H$_6$ and $n$-butane, the pools instead comprise coupled-exchange operators (CEOs) \cite{Ramoa2025CEO}. They preserve particle number and $S_z$; their pool and active-space specifications are given in Section~\ref{sec:computational-details}. The screening expressions below apply after a pool member has been represented as a Hermitian qubit generator $\hat G_\nu$.

This ansatz can be viewed as an adaptive truncation of a much larger wave-function expression: instead of including the entire pool at once, ADAPT retains only the operators that are selected during the iterative construction, so that typically $m \ll |\mathcal{P}|$. The algorithm therefore accelerates convergence with respect to \emph{ansatz size} by avoiding parameters that have little immediate effect on the energy, while still allowing the ansatz to grow systematically when more correlation is needed. This gain in compactness comes at the cost of repeatedly evaluating the operator-selection gradients.

At step $m$, the importance of a candidate pool operator $\hat{A}_{\nu}$ is measured by the first derivative of the energy obtained by appending an infinitesimal copy of that operator to the current ansatz,
\begin{equation}
\begin{aligned}
g_{\nu}^{(m)}
&=
\left|
\left.
\frac{\partial}{\partial \alpha}
\begin{gathered}
\langle \psi_A^{(m)} |
e^{-\alpha \hat{A}_{\nu}} \\
\hat{H}^{\mathrm{emb}}_A
e^{\alpha \hat{A}_{\nu}}
| \psi_A^{(m)} \rangle
\end{gathered}
\right|_{\alpha=0}
\right|
\\
&=
\left|
\langle \psi_A^{(m)} |
[\hat{H}^{\mathrm{emb}}_A,\hat{A}_{\nu}]
| \psi_A^{(m)} \rangle
\right|.
\end{aligned}
\end{equation}
Standard ADAPT chooses the operator with the largest gradient,
\begin{equation}
\nu^{\star} = \arg\max_{\nu \in \mathcal{P}} g_{\nu}^{(m)},
\end{equation}
appends $e^{\theta_{m+1}\hat{A}_{\nu^{\star}}}$ to the ansatz, initializes the new parameter to zero, and then reoptimizes \emph{all} selected parameters. The ADAPT iteration is terminated when the gradient norm satisfies
\begin{equation}
\|\mathbf{g}^{(m)}\|_2 < \varepsilon_{\mathrm{grad}},
\end{equation}
where $\mathbf{g}^{(m)}=(g_{\nu}^{(m)})_{\nu\in\mathcal{P}}$ is the vector of all pool gradients at iteration $m$, or when the change in energy between successive growth steps falls below a secondary threshold $\varepsilon_E$.

\subsection{FastAdaptVQE}

FastAdaptVQE is a statevector simulation backend that preserves the ADAPT ansatz construction, ranking rule, and convergence tests while replacing symbolic commutator construction by sparse linear algebra. Let $H^{\mathrm{emb}}_{A,\mathrm{sparse}}$ and $G_{\nu,\mathrm{sparse}}$ denote the sparse matrices of the Hamiltonian and the Hermitian qubit generator $\hat G_\nu=i\hat A_\nu$. At ADAPT iteration $m$, define
\begin{equation}
|h\rangle
=H^{\mathrm{emb}}_{A,\mathrm{sparse}}|\psi_A^{(m)}\rangle,
\qquad
|u_\nu\rangle
=G_{\nu,\mathrm{sparse}}|\psi_A^{(m)}\rangle,
\end{equation}
and
\begin{equation}
z_\nu^{(m)}
=\langle h|u_\nu\rangle
=\langle\psi_A^{(m)}|
\hat H_A^{\mathrm{emb}}\hat G_\nu
|\psi_A^{(m)}\rangle.
\end{equation}
Because both $\hat H_A^{\mathrm{emb}}$ and $\hat G_\nu$ are Hermitian,
\begin{equation}
\langle\psi_A^{(m)}|
\hat G_\nu\hat H_A^{\mathrm{emb}}
|\psi_A^{(m)}\rangle
=\left(z_\nu^{(m)}\right)^*.
\end{equation}
Using $\hat A_\nu=-i\hat G_\nu$, the anti-Hermitian derivative defined in the ADAPT subsection is therefore
\begin{align}
d_\nu^{(m)}
&=\langle\psi_A^{(m)}|
[\hat H_A^{\mathrm{emb}},\hat A_\nu]
|\psi_A^{(m)}\rangle \\
&=-i\left[z_\nu^{(m)}-\left(z_\nu^{(m)}\right)^*\right]
=2\,\operatorname{Im}z_\nu^{(m)}.
\end{align}
Equivalently, if $|a_\nu\rangle=A_{\nu,\mathrm{sparse}}|\psi_A^{(m)}\rangle$, then $d_\nu^{(m)}=2\,\operatorname{Re}\langle h|a_\nu\rangle$. The ranking quantity used by ADAPT is consequently
\begin{equation}
g_\nu^{(m)}
=|d_\nu^{(m)}|
=2\left|\operatorname{Im}\langle h|u_\nu\rangle\right|
=2\left|\operatorname{Re}\langle h|a_\nu\rangle\right|.
\end{equation}
The implementation evaluates the two commutator terms explicitly through sparse matrix--vector products. Circuit-backed libraries may report the signed raw gradient with the opposite overall generator convention; that sign does not affect the absolute ranking, convergence norm, or selected operator. FastAdaptVQE thus computes $|h\rangle$ once per ADAPT iteration and reuses it while screening the pool, without materializing any symbolic commutators.

After the gradients are ranked, FastAdaptVQE proceeds exactly as in the ADAPT algorithm above: the selected operator is appended to the ansatz, its new amplitude is initialized to zero, and all amplitudes are reoptimized in the updated ansatz before the next growth step. Thus, the method changes the representation used for candidate-gradient evaluation, but not the variational objective or the adaptive growth policy itself.

If $n_H$ is the number of Pauli terms in the Hamiltonian and $n_\nu$ is the number of Pauli terms in candidate operator $\nu$, then the sparse storage used by FastAdaptVQE scales as
\begin{equation}
M_{\mathrm{sparse}}
=
\mathcal{O}\!\left[
\left(
n_H + \sum_{\nu} n_\nu
\right) 2^n
\right].
\end{equation}
FastAdaptVQE therefore removes the symbolic-algebra bottleneck in ADAPT candidate-gradient evaluation, but it still requires storing sparse matrix representations of the Hamiltonian and the operator pool.

\subsection{MatrixFreeAdaptVQE}

MatrixFreeAdaptVQE is an alternative statevector simulation backend. It keeps the same ADAPT ansatz growth and FastAdaptVQE gradient expression but replaces stored sparse matrices by a compact representation of each Pauli term. For a Pauli word
\begin{equation}
\hat{P}_t = \sigma_{p_{n-1}} \otimes \cdots \otimes \sigma_{p_0},
\end{equation}
with each $\sigma_{p_j}\in\{I,X,Y,Z\}$, let $f_t$ be the bit mask whose 1-bits mark $X$ or $Y$, let $z_t$ mark $Y$ or $Z$, and let $n_{Y,t}$ be the number of $Y$ factors. For a computational-basis source index $s$,
\begin{equation}
\hat{P}_t |s\rangle
=
i^{n_{Y,t}}(-1)^{\operatorname{popcount}(s \wedge z_t)}
|s \oplus f_t\rangle,
\label{eq:pauli-source-action}
\end{equation}
where $\wedge$ and $\oplus$ denote bitwise AND and exclusive OR (XOR). For $\hat O=\sum_t c_t\hat P_t$, the amplitude at output index $b$ is therefore
\begin{equation}
[\hat{O}\psi]_b
=
\sum_t c_t(-i)^{n_{Y,t}}
(-1)^{\operatorname{popcount}(b \wedge z_t)}
\psi_{b \oplus f_t}.
\label{eq:pauli-output-action}
\end{equation}
Equation~\eqref{eq:pauli-output-action} follows from Eq.~\eqref{eq:pauli-source-action} after setting $s=b\oplus f_t$ and using $\operatorname{popcount}(f_t\wedge z_t)=n_{Y,t}$. The implementation stores $\eta_t=c_t(-i)^{n_{Y,t}}$ together with $f_t$ and $z_t$, so it can apply $\hat O$ without forming its matrix.

The stored objects are the current statevector, a cached array of basis indices $b=0,\ldots,2^n-1$, and the tuples $(\eta_t,f_t,z_t)$ for every Hamiltonian and pool term. The matrix-free primitive forms the same $|h\rangle$ and $|u_\nu\rangle$ as FastAdaptVQE and evaluates the same $g_\nu^{(m)}$. Everything outside this routine, including the pool, ranking, convergence tests, ansatz growth, and inner reoptimization is unchanged.

Instead of the sparse scaling above, the matrix-free representation stores the statevector plus compact term metadata,
\begin{equation}
M_{\mathrm{mf}}
=
\mathcal{O}\!\left(2^n+n_H+\sum_\nu n_\nu\right).
\end{equation}
This removes the sparse-operator storage wall, although statevector simulation still scales exponentially with $n$.

\subsection{Lookahead Operator Selection for ADAPT-VQE}

The lookahead selector replaces a greedy top-1 choice by a one-step energy test over a small candidate set. Let $\mathcal{S}^{(m)}$ be the candidates sorted by decreasing $|g_\nu^{(m)}|$, and let $\mathcal{T}_k^{(m)}$ contain the first $k$ members above the ADAPT threshold. Two policies are evaluated. For H$_4$ and F$_2$, lookahead is triggered only after the selected ansatz shows a recurrence pattern. With $W^{(m)}=\{\nu_{m-w},\ldots,\nu_{m-1}\}$, activation requires $m\ge m_{\min}$, the top-gradient operator to occur in $W^{(m)}$, and either $ |\mathcal{T}_k^{(m)}\cap W^{(m)}|\ge r_{\min} $ or a member of $\mathcal{T}_k^{(m)}$ that would create a repeated block. Once triggered within a fragment solve, lookahead remains active. The shortlist $\mathcal{C}_k^{(m)}\subseteq\mathcal{T}_k^{(m)}$ removes a candidate if appending its pool index would produce identical consecutive indices or a consecutive repeated block
\begin{equation}
(\nu_j,\ldots,\nu_{j+L-1},\nu_j,\ldots,\nu_{j+L-1}),\qquad L\ge2.
\end{equation}
This index-pattern test is a search heuristic, not a statement that the corresponding operator products are algebraically redundant. For H$_6$ and $n$-butane, an always-on policy applies lookahead at every growth step to up to the top $k$ candidates and uses neither the recurrence trigger nor the cyclicity filter.

For each $\hat A_\nu\in\mathcal C_k^{(m)}$, define the trial state
\begin{equation}
|\psi_{\nu,\mathrm{trial}}^{(m+1)}(\boldsymbol\varphi,\alpha)\rangle
=e^{\alpha\hat A_\nu}
e^{\varphi_m\hat A_{\nu_m}}\cdots e^{\varphi_1\hat A_{\nu_1}}
|\phi_A^{\mathrm{ref}}\rangle
\end{equation}
and its locally reoptimized score
\begin{equation}
\widetilde E_\nu^{(m+1)}=
\min_{\boldsymbol\varphi,\alpha}
\langle\psi_{\nu,\mathrm{trial}}^{(m+1)}|
\hat H_A^{\mathrm{emb}}|
\psi_{\nu,\mathrm{trial}}^{(m+1)}\rangle.
\end{equation}
All existing amplitudes and the appended amplitude are reoptimized. The initial point is the current optimized vector with one new zero amplitude; a later trial of the same candidate may use its most recent trial value.

The appended operator is
\begin{equation}
\nu^\star=\arg\min_{\nu\in\mathcal C_k^{(m)}}\widetilde E_\nu^{(m+1)}.
\end{equation}
Energies within $10^{-12}$~Ha are tied, in which case the better raw-gradient rank is selected. The winning trial parameters seed the subsequent full inner reoptimization. Lookahead therefore changes a package of selection operations---shortlisting, full trial reoptimization, winner selection, and parameter seeding---rather than the raw ranking rule alone. The triggered policy adds cost only after activation; the always-on policy requires up to $k$ trial optimizations at every growth step. A local ranking reversal can repair a greedy trajectory, but the procedure does not guarantee a better final ansatz.

\subsection{Computational details}
\label{sec:computational-details}

All molecular calculations use the STO-3G basis, restricted Hartree--Fock orbitals, Jordan--Wigner mapping, and noiseless classical expectation values. BE uses the atom-based fragmentation described above, an outer tolerance of $10^{-6}$, and at most 50 outer iterations.

For the linear H$_4$ chain, the nearest-neighbor spacing is 1.0~\AA. BE2 produces the overlapping atom sets $[\mathrm{H}_1,\mathrm{H}_2,\mathrm{H}_3]$ and $[\mathrm{H}_2,\mathrm{H}_3,\mathrm{H}_4]$. Each embedded problem has four spatial orbitals and four electrons (8 qubits). For F$_2$ at 1.42~\AA, BE1 produces one fragment per fluorine atom; each embedded problem has six spatial orbitals and ten electrons (12 qubits), and the outer condition is only the chemical-potential electron-count constraint. Both systems use the L\"owdin local basis, $\tau_{\mathrm{bath}}=10^{-10}$, and the non-generalized singles-and-doubles pool. Their ADAPT thresholds are $\varepsilon_{\mathrm{grad}}=10^{-3}$ and $\varepsilon_E=10^{-5}$; greedy and triggered-lookahead runs are capped at 20 growth steps. The main SLSQP reoptimizations allow 100 iterations per stage with function tolerances $10^{-3}$, $10^{-4}$, and $10^{-6}$; the lookahead trials allow 40 iterations with tolerance $10^{-10}$. The trigger parameters are $k=5$, $m_{\min}=5$, $w=5$, and $r_{\min}=3$.

The F$_2$ pair observable reported below is the chemists'-convention spatial two-particle RDM element
\begin{equation}
p_\sigma=\langle n_{2\alpha}n_{2\beta}\rangle=\Gamma_{2222},
\end{equation}
where spatial orbital 2 is the bonding $2\sigma_g$ orbital and orbital 5 is the target $2\sigma_u^*$ virtual orbital used in the pair-excitation analysis.

For linear H$_6$, the uniform H--H spacing is 1.0, 1.5, or 2.0~\AA. BE2 produces four overlapping three-atom motifs: $[\mathrm{H}_1,\mathrm{H}_2,\mathrm{H}_3]$, $[\mathrm{H}_2,\mathrm{H}_3,\mathrm{H}_4]$, $[\mathrm{H}_3,\mathrm{H}_4,\mathrm{H}_5]$, and $[\mathrm{H}_4,\mathrm{H}_5,\mathrm{H}_6]$. Each embedded problem has six spatial orbitals and six electrons (12 qubits). H$_6$ ADAPT runs are capped at 50 growth steps. The lookahead tests up to the top five candidates at every step and reoptimizes with SLSQP (200 iterations, tolerance $10^{-10}$).

For $n$-butane, the source geometry is the pinned PubChem compound identifier (CID) 7843 three-dimensional conformer 00001EA3\allowbreak 00000001, validated as a neutral-singlet anti conformer with a $178.8496819^\circ$ C1--C2--C3--C4 dihedral. The scan rigidly translates C3, C4, and their attached hydrogens along the C2-to-C3 axis so that only the C2--C3 distance changes to 1.30, 1.54, or 2.10~\AA. BE2 uses two locked carbon-centered motifs, C1--C2--C3 and C2--C3--C4. The lookahead protocol matches H$_6$.

The 18-qubit screening benchmark and the 20-qubit memory stress test were run on an Amazon Web Services (AWS) c6i.4xlarge instance with 16 virtual central processing units (vCPUs), 32~GiB of random-access memory (RAM), and four OpenMP threads. Screening wall time was measured for one full operator-pool gradient pass. The legacy symbolic implementation did not finish that pass before it was terminated, so the performance comparison is reported as a lower bound.

The software environment was Python 3.12.11, Qiskit 2.2.3, Qiskit Aer 0.17.2, Qiskit Algorithms 0.4.0, Qiskit Nature 0.7.2, and PySCF 2.11.0.

\section{Results}

Except in Section 3.3, the reference in this section is FCI-BE: each embedded fragment is solved by full configuration interaction within the same fragment, bath, active space, and outer BE procedure used for the corresponding variational calculation.

\subsection{Backend validation and computational performance}

We first tested whether the two accelerated statevector backends preserve the reference ADAPT-VQE result. Table~\ref{tab:exact_validation} compares both FastAdaptVQE and MatrixFreeAdaptVQE with the same reference ADAPT-VQE calculation. FastAdaptVQE agrees within 0.0032~$\mu$Ha for H$_4$ and 2.7786~$\mu$Ha for F$_2$; MatrixFreeAdaptVQE agrees within 0.0005 and 1.2936~$\mu$Ha, respectively. These differences are negligible on the energy scales discussed below and show that the implementation strategies preserve the optimized noiseless-statevector results in these controls.

The performance benefits address different bottlenecks. In the 18-qubit operator-screening benchmark, the legacy symbolic-commutator path (which means without using FastAdaptVQE) was stopped after more than 42~h without completing a single pool-gradient pass, whereas the FastAdaptVQE pass took 1.9~s. In the 20-qubit memory stress test, sparse conversion (which means without using MatrixFreeAdaptVQE) reached approximately 107~GB of virtual memory and 30~GB resident memory before failing on the 32~GiB node; the MatrixFreeAdaptVQE process remained near 662~MB while completing the same gradient evaluation.

\begin{table}[htbp]
\centering
\caption{Noiseless-statevector validation of the accelerated ADAPT
implementations. Both error columns use the same reference ADAPT-VQE
bootstrap-embedding energy. Energies are in Hartree.}
\label{tab:exact_validation}

\resizebox{\linewidth}{!}{\begin{tabular}{lccccc}
\hline
System & Reference ADAPT-VQE & FastAdaptVQE & $|\Delta E_{\mathrm{fast}}|$ ($\mu$Ha)
& MatrixFreeAdaptVQE & $|\Delta E_{\mathrm{mf}}|$ ($\mu$Ha) \\
\hline
H$_4$ & -2.1661448670 & -2.1661448638 & 0.0032
& -2.1661448675 & 0.0005 \\
F$_2$ & -195.9728711800 & -195.9728739586 & 2.7786
& -195.9728724736 & 1.2936 \\
\hline
\end{tabular}
}
\end{table}

\subsection{H$_4$: fragment-solver compatibility control}

The H$_4$ BE2 calculation tests whether a variational fragment solver can be inserted into the self-consistent embedding loop without disrupting the embedding. Table~\ref{tab:h4_exact} shows that UCCSD VQE-BE and ADAPT-VQE-BE remain within chemical accuracy of FCI-BE, with errors of +0.4197 and +0.2426~mHa, respectively. The lookahead run provides an important negative control. Its activation criterion was never satisfied in either fragment solve, so it followed the greedy trajectory and returned the same final energy. Both calculations ended after the first outer VQE-BE iteration with a BE2 matching residual of $1.401426\times10^{-7}$.

For Tables~\ref{tab:h4_exact} and \ref{tab:f2_exact}, define \(\Delta E_{\mathrm{BE}}=1000(E_{\mathrm{BE}}-E_{\mathrm{FCI-BE}})\) in mHa. The correlation recovered percentage is measured relative to the corresponding FCI-BE correlation energy.

\begin{table}[htbp]
\centering
\caption{Noiseless-statevector H$_4$ compatibility control. The solver
reference is $E_{\mathrm{FCI-BE}}=-2.1663874445$~Ha.}
\label{tab:h4_exact}
\small
\setlength{\tabcolsep}{7pt}
\begin{tabular}{lccc}
\hline
Method & $E_{\mathrm{BE}}$ (Ha) & $\Delta E_{\mathrm{BE}}$ (mHa) & Correlation recovered (\%) \\
\hline
HF & -2.0985459370 & +67.8415 & 0.00 \\
UCCSD VQE-BE & -2.1659677386 & +0.4197 & 99.38 \\
Greedy ADAPT-VQE-BE & -2.1661448672 & \textbf{+0.2426} & \textbf{99.64} \\
Triggered-lookahead ADAPT-VQE-BE & -2.1661448672 & \textbf{+0.2426} & \textbf{99.64} \\
\hline
\end{tabular}
\end{table}

\subsection{H$_6$: matching consistency under different geometries}
\begin{figure}[htbp]
\centering
\includegraphics[width=1.0\linewidth]{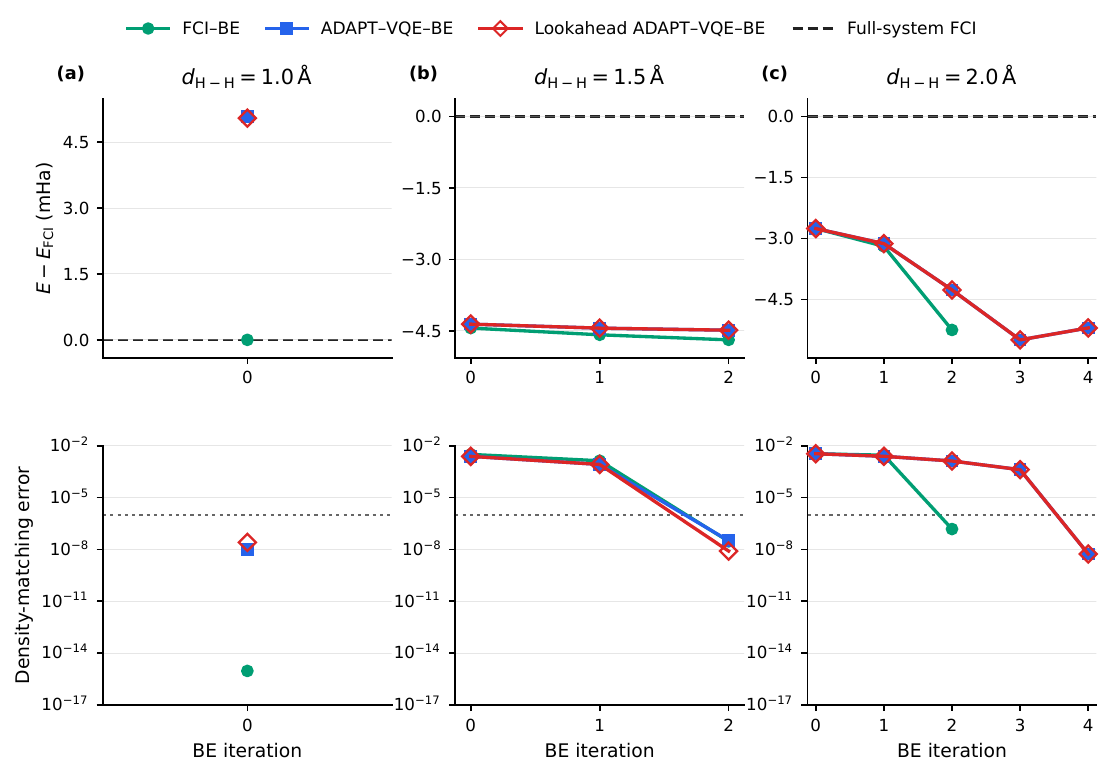}
\caption{FCI-BE, greedy ADAPT-VQE-BE, and lookahead ADAPT-VQE-BE convergence
for linear H$_6$ at
$d_{\mathrm{H-H}}=$ (a) 1.0, (b) 1.5, and (c) 2.0~\AA, where
$d_{\mathrm{H-H}}$ denotes the uniform nearest-neighbor H--H separation. The top row shows the signed BE energy
relative to full-system FCI; the dashed line marks zero. The bottom row shows the density-matching error. The UCCSD VQE-BE results are not shown as the error in energy relative to FCI
in (a) was +72.087113 mHa. }
\label{fig:h6_qbe_production}
\end{figure}

To test the coupled VQE--QBE loop beyond the H$_4$ control, BE2 was applied to a linear H$_6$ chain in STO-3G at uniform nearest-neighbor H--H separations of $d_{\mathrm{H-H}}=1.0$, 1.5, and 2.0~\AA{}. Figure~\ref{fig:h6_qbe_production} shows that both greedy and lookahead VQE-BE converged the density-matching error below $3.3\times10^{-8}$ in all three calculations. Note that because the assembled BE energy is not globally variational, FCI-BE and VQE-BE may lie below this reference. Relative to the corresponding FCI-BE controls, the final greedy errors were +5.084, +0.202, and +0.050~mHa, respectively. At 1.0~\AA, lookahead reduced the error by only 0.038~mHa, although 92 of the 200 operator-selection events were non-greedy. It matched greedy to numerical precision at 1.5 and 2.0~\AA, so it did not materially improve H$_6$ energy accuracy. Stronger short-range orbital overlap and interfragment coupling could make the 1.0~\AA{} assembled energy more sensitive to fragment-state errors. Within each resolved fragmentation, the results show that density convergence alone does not guarantee energy accuracy.

\subsection{F$_2$: lookahead resolves greedy ADAPT-VQE-BE plateau}

We then tested our ADAPT-VQE-BE algorithm on a chemically realistic system, F$_2$. With exact expectation values and the greedy ADAPT selector, the BE total energy stalls at $-195.9728739586$~Ha, +71.3708~mHa above FCI-BE, while the final fragment energy remains +188.7915~mHa above the FCI fragment reference $-103.2502698403$~Ha. At the same time, the bonding $\sigma$-pair density collapses to 0.00739, far below the FCI-BE value of 0.11444. Table~\ref{tab:f2_exact} shows that replacing only the way operators are selected resolves this failure as the lookahead selector recovers a final BE energy of $-196.0442411891$~Ha, only +0.00355~mHa above FCI-BE, together with a final fragment energy +0.07329~mHa from the FCI fragment and a restored bonding $\sigma$-pair density of 0.11445.

\begin{figure}[htbp]
\centering
\includegraphics[width=1.0\linewidth]{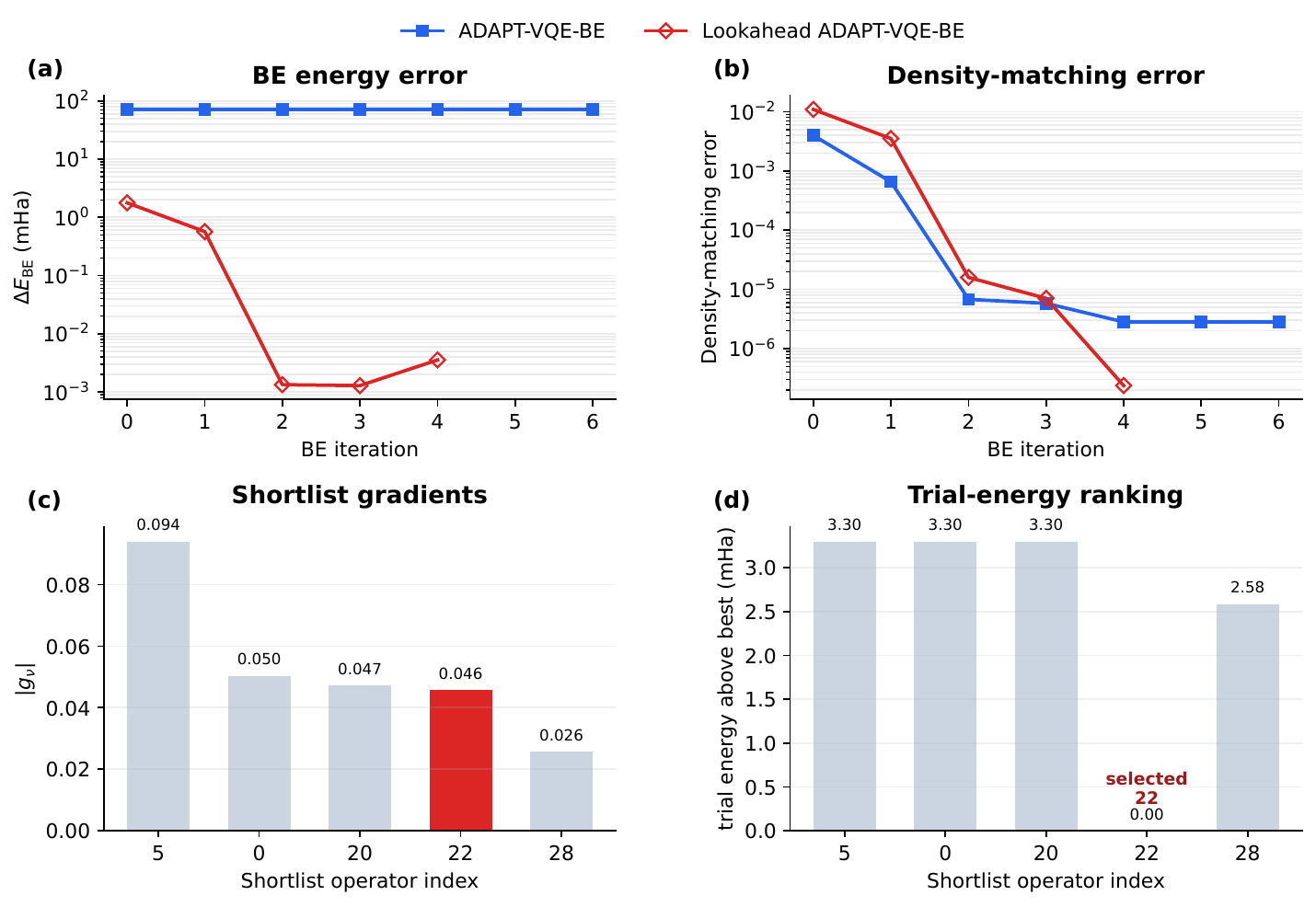}
\caption{F$_2$ lookahead rescue in the statevector regime. (a)
$\Delta E_{\mathrm{BE}}$ versus BE iteration and (b) density-matching error
versus BE iteration for greedy ADAPT-VQE and the lookahead-enabled run. Panels
(c) and (d) show the first lookahead decision from the production run for the
non-cyclic shortlisted operators: (c) gives the raw gradient magnitudes, and
(d) gives the trial energy offsets relative to the best shortlisted candidate,
both using shortlist operator indices on the x-axis. The overall raw rank-2
candidate (operator~16) is omitted because its inclusion would have produced
an operator cycle. Operator~22 is fifth overall by raw
gradient magnitude, but it gives the lowest post-probe energy among the
shortlisted non-cyclic candidates and is therefore selected.}
\label{fig:f2_lookahead_rescue}
\end{figure}

The decisive point is that the missing move was already present in the operator pool. At the first selector event of the lookahead run, operator~22 (\texttt{2a,2b->5a,5b}) ranked only fifth by raw gradient magnitude, yet its locally reoptimized trial energy was 3.30~mHa lower than the rank-1 candidate and 2.58~mHa lower than the next-best shortlisted alternative. Figure~\ref{fig:f2_lookahead_rescue} shows both the corrected BE trajectory and this ranking reversal directly. This result identifies operator selection, rather than the operator pool, as the source of the F$_2$ failure.

\begin{table}[htbp]
\centering
\caption{Exact-state F$_2$ benchmark. The FCI-BE reference is
$E_{\mathrm{BE}}^{\mathrm{FCI-BE}}=-196.0442447354$~Ha, and the FCI bonding
$\sigma$-pair density is 0.11444.}
\label{tab:f2_exact}
\small
\resizebox{\linewidth}{!}{\begin{tabular}{lcccc}
\hline
Case & $E_{\mathrm{BE}}$ (Ha) & $\Delta E_{\mathrm{BE}}$ (mHa) & Correlation recovered (\%) & Bonding $\sigma$ pair \\
\hline
HF & -195.9652604114 & +78.9843 & 0.00 & --- \\
Greedy ADAPT-VQE-BE & -195.9728739586 & +71.3708 & 9.64 & 0.00739 \\
Lookahead ADAPT-VQE-BE & -196.0442411891 & \textbf{+0.00355} & \textbf{100.00} & 0.11445 \\
\hline
\end{tabular}
}
\end{table}
\bigskip

\subsubsection{Why Operator 22 Was Chosen}
To identify the mechanism behind this energy improvement in F$_2$, we examined why lookahead selected operator~22. To make the notation explicit, define for each occupied orbital $r \in \{0,1,2,3,4\}$ the same-orbital pair-excitation family
\begin{equation}
\mathcal{P}_r = \{r\alpha \rightarrow 5\alpha,\ r\beta \rightarrow 5\beta,\ r\alpha,r\beta \rightarrow 5\alpha,5\beta\},
\end{equation}
where orbital 5 is the common target virtual orbital in this F$_2$ BE1 pool. These $\mathcal{P}_r$ symbols are local bookkeeping labels for the five possible same-orbital pair channels into orbital~5. The corresponding occupied orbitals are $1\sigma_g$, $1\sigma_u^\ast$, $2\sigma_g$, $2\pi_{u,x}$, and $2\pi_{u,y}$ for $r=0,1,2,3,4$, respectively.

On the exact captured selector state, the channels fall into three distinct classes. A channel is \emph{completed} if both singles and the paired double are already present in the ansatz prefix; this is the case for $\mathcal{P}_1$. A channel is \emph{activated} if the two singles are already present but the paired double is still missing; this is the case for $\mathcal{P}_0$ and $\mathcal{P}_2$. A channel is \emph{dormant} if neither the singles nor the paired double have yet been completed; this is the case for $\mathcal{P}_3$ and $\mathcal{P}_4$.

Operator~22 is therefore the completion move of the valence $2\sigma_g^2 \rightarrow 2\sigma_u^{\ast\,2}$ pair channel $\mathcal{P}_2$, which is the missing correlation sector in this F$_2$ failure mode. By contrast, operator~10 completes the core $1\sigma_g^2 \rightarrow 2\sigma_u^{\ast\,2}$ channel $\mathcal{P}_0$ and is 2.54~mHa worse after local reoptimization even though its singles are already activated. The dormant $\pi \rightarrow \sigma^\ast$ pair completions 28 and 34, from $\mathcal{P}_3$ and $\mathcal{P}_4$, respectively, outperform operator 10, but each remains 1.78~mHa worse than 22. The lookahead selector therefore succeeds because it identifies the dominant incomplete activated pair completion.

\begin{table}[htbp]
\centering

\caption{Selector-state analysis for the first decisive F$_2$ lookahead event.
Here $\mathcal{P}_r$ denotes the same-orbital pair family built from occupied orbital
$r$ and the common target virtual orbital 5, i.e.\
$\mathcal{P}_r=\{r\alpha \rightarrow 5\alpha,\ r\beta \rightarrow 5\beta,\ r\alpha,r\beta \rightarrow 5\alpha,5\beta\}$.
The ranking is evaluated on the exact captured plateau state.}
\label{tab:f2_pair_channels}
\small
\resizebox{\linewidth}{!}{\begin{tabular}{lcccccc}
\hline
Family & Occupied orbital & Double & Status at activation & Raw gradient rank & Trial-energy rank & Gap to best trial energy (mHa) \\
\hline
$\mathcal{P}_0$ & $1\sigma_g$ & 10 & activated & 12 & 4 & +2.537 \\
$\mathcal{P}_1$ & $1\sigma_u^\ast$ & 16 & completed & 2 & --- & --- \\
$\mathcal{P}_2$ & $2\sigma_g$ & 22 & activated & 5 & \textbf{1} & \textbf{0.000} \\
$\mathcal{P}_3$ & $2\pi_{u,x}$ & 28 & dormant & 6 & 2 & +1.781 \\
$\mathcal{P}_4$ & $2\pi_{u,y}$ & 34 & dormant & 7 & 3 & +1.781 \\
\hline
\end{tabular}
}
\end{table}

\subsection{C$_4$H$_{10}$: larger system validation of ADAPT-VQE-BE}

\begin{figure}[htbp]
\centering
\includegraphics[width=1.0\linewidth]{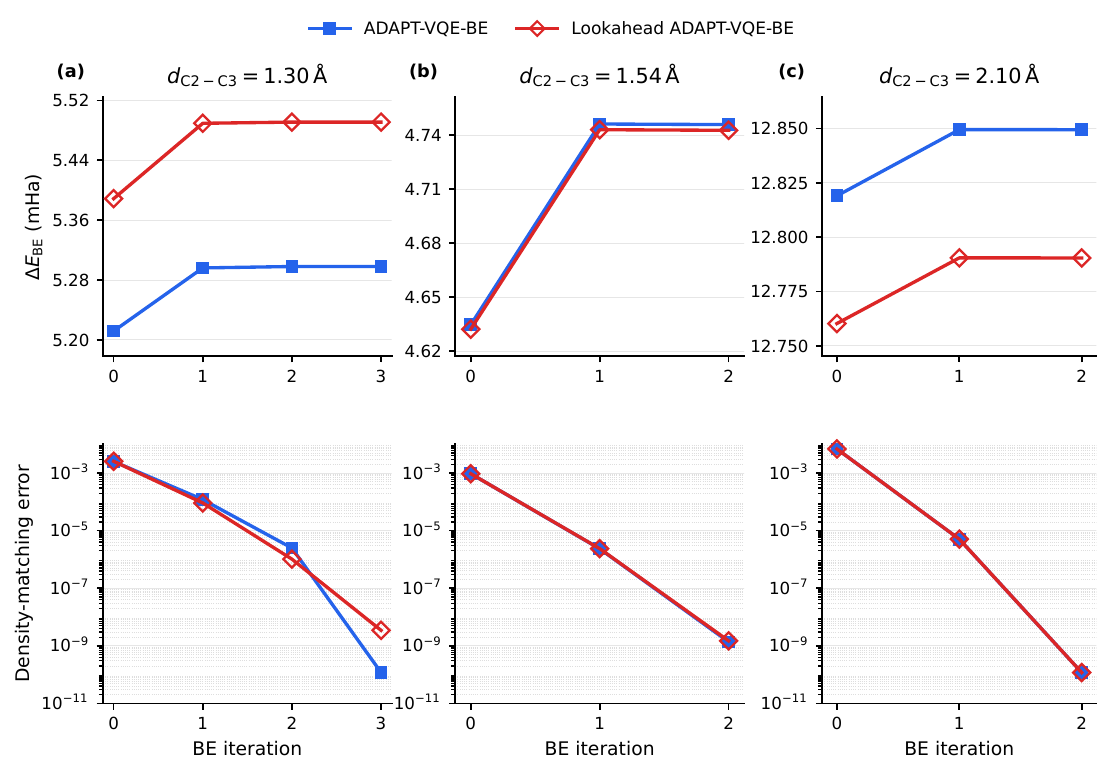}
\caption{Greedy and lookahead ADAPT-VQE-BE convergence for
$n$-butane. Columns (a)--(c) correspond to
$d_{\mathrm{C2-C3}}=$ 1.30, 1.54, and 2.10~\AA. The top row shows
$\Delta E_{\mathrm{BE}}=|E_{\mathrm{BE}}-E_{\mathrm{FCI-BE}}|$; the bottom row
shows the density-matching residual per iteration.}
\label{fig:c4h10_w12_qbe}
\end{figure}

We then applied two-fragment BE2 to $n$-butane. Figure~\ref{fig:c4h10_w12_qbe} shows results for three distances between the second and third carbons. At $d_{\mathrm{C2-C3}}=1.30$, 1.54, and 2.10~\AA, the final greedy (lookahead) errors relative to FCI-BE were 5.298 (5.291), 4.746 (4.743), and 12.849 (12.790)~mHa, respectively. Both selectors converged the density-matching residual below $3.5\times10^{-9}$, showing that the coupled VQE-BE procedure remains stable and reaches this tight density consistency even in this larger molecular benchmark. Lookahead reduced the error by at most 0.059~mHa, so it did not provide a material accuracy gain in this system.

\section{Discussion}

Taken together, these benchmarks show that VQE--QBE depends on the quality of the fragment states and their reduced density matrices (RDMs). H$_4$ shows that a sufficiently converged VQE solver combined with BE can reproduce the FCI-BE solution. H$_6$ and $n$-butane demonstrate that the embedding loop remains stable across bond geometries and larger fragments even when fragment-solver errors persist in the assembled energy. F$_2$ gives the sharpest failure mode, in which the outer loop cannot repair an incorrect fragment state and RDM, showing the potential of the lookahead strategy. Because finite sampling and device noise can further perturb these quantities, a practical operator-selection procedure must remain optimizable as the embedded Hamiltonians change while returning RDMs accurate enough both to satisfy the embedding constraints and to evaluate the assembled energy
\cite{Welborn2016,Ye2019,Ye2020,Liu2023QBE}.

The three algorithmic contributions address distinct inner-solver bottlenecks. FastAdaptVQE removes symbolic-commutator construction, MatrixFreeAdaptVQE avoids storing large sparse operators, and lookahead revises operator selection when a local gradient ranking is misleading. The accelerated backends preserve the reference exact-state solutions, whereas limited trial optimization over a shortlist corrects the specific greedy failure observed for F$_2$. In the present benchmarks, lookahead is most effective in the single case in which the greedy procedure becomes trapped in a local minimum, and the specific parameters used here do not generalize to the other systems studied. We hypothesize that more carefully tuned lookahead parameters, including the shortlist size, SLSQP iteration budget, and activation thresholds, could yield larger energy improvements relative to greedy ADAPT-VQE-BE.

FastAdaptVQE and MatrixFreeAdaptVQE substantially improve classical statevector simulation, but they do not change its exponential scaling or reduce hardware measurement costs. QBE remains advantageous only while embedded fragments stay much smaller than the full system. If longer range correlation increases, the fragments, operator pools, RDM estimation, and self-consistency will all become more expensive \cite{Tilly2022,Ye2020,Tran2024,Meitei2023Periodic,Cho2025QuEmb,Liu2023QBE}.

The present calculations use noiseless expectation values and a small STO-3G benchmark set. H$_6$ and $n$-butane add overlapping-fragment and geometry tests, but do not address finite-shot noise or larger active spaces. F$_2$ uses BE1 and does not test spatial overlap. The next studies should incorporate noise and larger basis sets, together with systems that exhibit more delocalized correlation. These extensions will test whether the same RDM-consistency and operator-selection requirements persist outside the noiseless statevector regime.

\section{Conclusion}

This work developed and evaluated a noiseless statevector VQE--QBE workflow. FastAdaptVQE removed symbolic-commutator overhead, while MatrixFreeAdaptVQE avoided the sparse-matrix memory barrier. A limited lookahead procedure addressed a separate ansatz-construction problem by re-evaluating promising ADAPT-VQE operators after local optimization.

The molecular benchmarks show that QBE accuracy depends on both the embedding conditions and the fragment solver. For H$_4$, ADAPT-VQE-BE reproduced the FCI-BE reference within chemical accuracy. For H$_6$, density matching converged at every geometry; VQE-BE reached chemical accuracy relative to FCI-BE at H--H spacings of 1.5 and 2.0~\AA. For $n$-butane, the density errors all fell below $3.5\times10^{-9}$, but final FCI-BE errors remained outside chemical accuracy. In the F$_2$ case, lookahead reduced the greedy FCI-BE error from 71.371 to 0.00355~mHa. In all cases, the lookahead method either preserved greedy ADAPT-VQE-BE energy accuracy or improved it.

More broadly, this work establishes an algorithmic foundation for extending VQE-based bootstrap embedding to larger molecules and related fragment-based quantum algorithms. Although the present study was carried out on modest benchmark systems, it defines a clear path forward for near-term implementations: combine qubit-efficient embedding with scalable adaptive solvers, more strategic ansatz-growth policies, and hardware-aware implementations that preserve these gains under noise and finite sampling. As quantum architectures move toward fault tolerance, the same embedding perspective may also prove useful for deeper algorithms such as quantum phase estimation, where reducing fragment Hamiltonian sizes could lower circuit resource requirements \cite{AspuruGuzik2005,Cao2019,McArdle2020,Bauer2020}. In this sense, the present results move quantum bootstrap embedding from a conceptual possibility toward a practical framework for molecular and materials simulation on near-term quantum devices.

\section{Software and Data Availability}

The software and scripts required to reproduce the main results are available at \url{https://github.com/littlebullGit/QBEMol}.

\section{Acknowledgments}

We thank Professor Brenda Rubenstein for her guidance on the methodology and feedback on the manuscript. We thank Brown University's Center for Computation and Visualization for access to the Oscar high-performance computing (HPC) cluster.

\bibliography{mybib}

@article{Preskill2018,
  author = {Preskill, John},
  title = {Quantum Computing in the {NISQ} Era and Beyond},
  journal = {Quantum},
  year = {2018},
  volume = {2},
  pages = {79},
  doi = {10.22331/q-2018-08-06-79},
  url = {https://doi.org/10.22331/q-2018-08-06-79}
}

@article{Cao2019,
  author = {Cao, Yudong and Romero, Jonathan and Olson, Jonathan P. and Degroote, Matthias and Johnson, Peter D. and Kieferov\'a, M\'aria and Kivlichan, Ian D. and Menke, Tim and Peropadre, Borja and Sawaya, Nicolas P. D. and Sim, Sukin and Veis, Libor and Aspuru-Guzik, Al\'an},
  title = {Quantum Chemistry in the Age of Quantum Computing},
  journal = {Chemical Reviews},
  year = {2019},
  volume = {119},
  number = {19},
  pages = {10856--10915},
  doi = {10.1021/acs.chemrev.8b00803},
  url = {https://doi.org/10.1021/acs.chemrev.8b00803}
}

@article{Cross2019,
  author = {Cross, Andrew W. and Bishop, Lev S. and Sheldon, Sarah and Nation, Paul D. and Gambetta, Jay M.},
  title = {Validating Quantum Computers Using Randomized Model Circuits},
  journal = {Physical Review A},
  year = {2019},
  volume = {100},
  number = {3},
  pages = {032328},
  doi = {10.1103/PhysRevA.100.032328},
  url = {https://doi.org/10.1103/PhysRevA.100.032328}
}

@article{Peruzzo2014,
  author = {Peruzzo, Alberto and McClean, Jarrod and Shadbolt, Peter and Yung, Man-Hong and Zhou, Xiao-Qi and Love, Peter J. and Aspuru-Guzik, Al\'an and O'Brien, Jeremy L.},
  title = {A Variational Eigenvalue Solver on a Photonic Quantum Processor},
  journal = {Nature Communications},
  year = {2014},
  volume = {5},
  pages = {4213},
  doi = {10.1038/ncomms5213},
  url = {https://doi.org/10.1038/ncomms5213}
}

@article{McClean2016,
  author = {McClean, Jarrod R. and Romero, Jonathan and Babbush, Ryan and Aspuru-Guzik, Al\'an},
  title = {The Theory of Variational Hybrid Quantum-Classical Algorithms},
  journal = {New Journal of Physics},
  year = {2016},
  volume = {18},
  number = {2},
  pages = {023023},
  doi = {10.1088/1367-2630/18/2/023023},
  url = {https://doi.org/10.1088/1367-2630/18/2/023023}
}

@article{Grimsley2019,
  author = {Grimsley, Harper R. and Economou, Sophia E. and Barnes, Edwin and Mayhall, Nicholas J.},
  title = {An Adaptive Variational Algorithm for Exact Molecular Simulations on a Quantum Computer},
  journal = {Nature Communications},
  year = {2019},
  volume = {10},
  number = {1},
  pages = {3007},
  doi = {10.1038/s41467-019-10988-2},
  url = {https://doi.org/10.1038/s41467-019-10988-2}
}

@article{Knizia2013,
  author = {Knizia, Gerald and Chan, Garnet Kin-Lic},
  title = {Density Matrix Embedding: A Strong-Coupling Quantum Embedding Theory},
  journal = {Journal of Chemical Theory and Computation},
  year = {2013},
  volume = {9},
  number = {3},
  pages = {1428--1432},
  doi = {10.1021/ct301044e},
  url = {https://doi.org/10.1021/ct301044e}
}

@article{Wouters2016,
  author = {Wouters, Sebastian and Jim\'enez-Hoyos, Carlos A. and Sun, Qiming and Chan, Garnet Kin-Lic},
  title = {A Practical Guide to Density Matrix Embedding Theory in Quantum Chemistry},
  journal = {Journal of Chemical Theory and Computation},
  year = {2016},
  volume = {12},
  number = {6},
  pages = {2706--2719},
  doi = {10.1021/acs.jctc.6b00316},
  url = {https://doi.org/10.1021/acs.jctc.6b00316}
}

@article{Welborn2016,
  author = {Welborn, Matthew and Tsuchimochi, Takashi and Van Voorhis, Troy},
  title = {Bootstrap Embedding: An Internally Consistent Fragment-Based Method},
  journal = {The Journal of Chemical Physics},
  year = {2016},
  volume = {145},
  number = {7},
  pages = {074102},
  doi = {10.1063/1.4960986},
  url = {https://doi.org/10.1063/1.4960986}
}

@article{Ye2019,
  author = {Ye, Hong-Zhou and Ricke, Nathan D. and Tran, Henry K. and Van Voorhis, Troy},
  title = {Bootstrap Embedding for Molecules},
  journal = {Journal of Chemical Theory and Computation},
  year = {2019},
  volume = {15},
  number = {8},
  pages = {4497--4506},
  doi = {10.1021/acs.jctc.9b00529},
  url = {https://doi.org/10.1021/acs.jctc.9b00529}
}

@article{Ye2019Atom,
  author = {Ye, Hong-Zhou and Van Voorhis, Troy},
  title = {Atom-Based Bootstrap Embedding for Molecules},
  journal = {The Journal of Physical Chemistry Letters},
  year = {2019},
  volume = {10},
  number = {20},
  pages = {6368--6374},
  doi = {10.1021/acs.jpclett.9b02479},
  url = {https://doi.org/10.1021/acs.jpclett.9b02479}
}

@article{Ye2020,
  author = {Ye, Hong-Zhou and Tran, Henry K. and Van Voorhis, Troy},
  title = {Bootstrap Embedding for Large Molecular Systems},
  journal = {Journal of Chemical Theory and Computation},
  year = {2020},
  volume = {16},
  number = {8},
  pages = {5035--5046},
  doi = {10.1021/acs.jctc.0c00438},
  url = {https://doi.org/10.1021/acs.jctc.0c00438}
}

@article{Tran2024,
  author = {Tran, Henry K. and Weisburn, Leah P. and Cho, Minsik and Weatherly, Shaun and Ye, Hong-Zhou and Van Voorhis, Troy},
  title = {Bootstrap Embedding for Molecules in Extended Basis Sets},
  journal = {Journal of Chemical Theory and Computation},
  year = {2024},
  volume = {20},
  number = {24},
  pages = {10912--10921},
  doi = {10.1021/acs.jctc.4c01267},
  url = {https://doi.org/10.1021/acs.jctc.4c01267}
}

@article{Meitei2023Periodic,
  author = {Meitei, Oinam Romesh and Van Voorhis, Troy},
  title = {Periodic Bootstrap Embedding},
  journal = {Journal of Chemical Theory and Computation},
  year = {2023},
  volume = {19},
  number = {11},
  pages = {3123--3130},
  doi = {10.1021/acs.jctc.3c00069},
  url = {https://doi.org/10.1021/acs.jctc.3c00069}
}

@article{Cho2025QuEmb,
  author = {Cho, Minsik and Meitei, Oinam Romesh and Weisburn, Leah P. and Weser, Oskar and Weatherly, Shaun and Alexiu, Alexandra and Hanscam, Rebecca and Tran, Henry K. and Ye, Hong-Zhou and Welborn, Matthew and Ricke, Nathan D. and Tsuchimochi, Takashi and Trofimov, Aleksandr and Orkhon, Temujin and Whelpley, Noah and Luo, Carina and Van Voorhis, Troy},
  title = {{QuEmb}: A Toolbox for Bootstrap Embedding Calculations of Molecular and Periodic Systems},
  journal = {The Journal of Physical Chemistry A},
  year = {2025},
  volume = {129},
  number = {28},
  pages = {6538--6551},
  doi = {10.1021/acs.jpca.5c02983},
  url = {https://doi.org/10.1021/acs.jpca.5c02983}
}

@article{Liu2023QBE,
  author = {Liu, Yuan and Meitei, Oinam R. and Chin, Zachary E. and Dutt, Arkopal and Tao, Max and Chuang, Isaac L. and Van Voorhis, Troy},
  title = {Bootstrap Embedding on a Quantum Computer},
  journal = {Journal of Chemical Theory and Computation},
  year = {2023},
  volume = {19},
  number = {8},
  pages = {2230--2247},
  doi = {10.1021/acs.jctc.3c00012},
  url = {https://doi.org/10.1021/acs.jctc.3c00012}
}

@article{Mineh2022,
  author = {Mineh, Lana and Montanaro, Ashley},
  title = {Solving the {Hubbard} Model Using Density Matrix Embedding Theory and the Variational Quantum Eigensolver},
  journal = {Physical Review B},
  year = {2022},
  volume = {105},
  number = {12},
  pages = {125117},
  doi = {10.1103/PhysRevB.105.125117},
  url = {https://doi.org/10.1103/PhysRevB.105.125117}
}

@article{Li2022,
  author = {Li, Weitang and Huang, Zigeng and Cao, Changsu and Huang, Yifei and Shuai, Zhigang and Sun, Xiaoming and Sun, Jinzhao and Yuan, Xiao and Lv, Dingshun},
  title = {Toward Practical Quantum Embedding Simulation of Realistic Chemical Systems on Near-Term Quantum Computers},
  journal = {Chemical Science},
  year = {2022},
  volume = {13},
  number = {31},
  pages = {8953--8962},
  doi = {10.1039/D2SC01492K},
  url = {https://doi.org/10.1039/D2SC01492K}
}

@article{Bierman2026,
  author = {Bierman, Joel and Liu, Yuan},
  title = {Towards Utility-Scale Electronic Structure with Sample-Based Quantum Bootstrap Embedding},
  journal = {Digital Discovery},
  year = {2026},
  volume = {5},
  number = {2},
  pages = {945--956},
  doi = {10.1039/D5DD00416K},
  url = {https://doi.org/10.1039/D5DD00416K}
}

@techreport{Kraft1988SQP,
  author      = {Kraft, Dieter},
  title       = {A Software Package for Sequential Quadratic Programming},
  institution = {Deutsche Forschungs- und Versuchsanstalt f\"ur Luft- und Raumfahrt (DFVLR), Institut f\"ur Dynamik der Flugsysteme},
  number      = {DFVLR-FB 88-28},
  address     = {Oberpfaffenhofen, Germany},
  month       = jul,
  year        = {1988}
}

@article{Kraft1994TOMP,
  author  = {Kraft, Dieter},
  title   = {Algorithm 733: {TOMP}--{Fortran} modules for optimal control calculations},
  journal = {ACM Transactions on Mathematical Software},
  volume  = {20},
  number  = {3},
  pages   = {262--281},
  month   = sep,
  year    = {1994},
  doi     = {10.1145/192115.192124}
}

@article{AspuruGuzik2005,
  author = {Aspuru-Guzik, Al{\'a}n and Dutoi, Anthony D. and Love, Peter J. and Head-Gordon, Martin},
  title = {Simulated Quantum Computation of Molecular Energies},
  journal = {Science},
  volume = {309},
  number = {5741},
  pages = {1704--1707},
  year = {2005},
  doi = {10.1126/science.1113479}
}

@article{McArdle2020,
  author = {McArdle, Sam and Endo, Suguru and Aspuru-Guzik, Al{\'a}n and Benjamin, Simon C. and Yuan, Xiao},
  title = {Quantum computational chemistry},
  journal = {Reviews of Modern Physics},
  volume = {92},
  pages = {015003},
  year = {2020},
  doi = {10.1103/RevModPhys.92.015003}
}

@article{Bauer2020,
  author = {Bauer, Bela and Bravyi, Sergey and Motta, Mario and Chan, Garnet Kin-Lic},
  title = {Quantum Algorithms for Quantum Chemistry and Quantum Materials Science},
  journal = {Chemical Reviews},
  volume = {120},
  number = {22},
  pages = {12685--12717},
  year = {2020},
  doi = {10.1021/acs.chemrev.9b00829}
}

@article{Bharti2022,
  author = {Bharti, Kishor and Cervera-Lierta, Alba and Kyaw, Thi Ha and Haug, Tobias and Alperin-Lea, Sumner and Anand, Abhinav and Degroote, Matthias and Heimonen, Hermanni and Kottmann, Jakob S. and Menke, Tim and Mok, Wai-Keong and Sim, Sukin and Kwek, Leong-Chuan and Aspuru-Guzik, Al{\'a}n},
  title = {Noisy intermediate-scale quantum algorithms},
  journal = {Reviews of Modern Physics},
  volume = {94},
  pages = {015004},
  year = {2022},
  doi = {10.1103/RevModPhys.94.015004}
}

@article{Cerezo2021,
  author = {Cerezo, M. and Arrasmith, A. and Babbush, R. and Benjamin, S. C. and Endo, S. and Fujii, K. and McClean, J. R. and Mitarai, K. and Yuan, X. and Cincio, L. and Coles, P. J.},
  title = {Variational quantum algorithms},
  journal = {Nature Reviews Physics},
  volume = {3},
  number = {9},
  pages = {625--644},
  year = {2021},
  doi = {10.1038/s42254-021-00348-9}
}

@article{Endo2021,
  author = {Endo, Suguru and Cai, Zhenyu and Benjamin, Simon C. and Yuan, Xiao},
  title = {Hybrid Quantum-Classical Algorithms and Quantum Error Mitigation},
  journal = {Journal of the Physical Society of Japan},
  volume = {90},
  number = {3},
  pages = {032001},
  year = {2021},
  doi = {10.7566/JPSJ.90.032001}
}

@article{Tilly2022,
  author = {Tilly, Jules and Chen, Hongxiang and Cao, Shuxiang and Picozzi, Dario and Setia, Kanav and Li, Ying and Grant, Edward and Wossnig, Leonard and Rungger, Ivan and Booth, George H. and Tennyson, Jonathan},
  title = {The Variational Quantum Eigensolver: A review of methods and best practices},
  journal = {Physics Reports},
  volume = {986},
  pages = {1--128},
  year = {2022},
  doi = {10.1016/j.physrep.2022.08.003}
}

@article{Kandala2017,
  author = {Kandala, Abhinav and Mezzacapo, Antonio and Temme, Kristan and Takita, Maika and Brink, Markus and Chow, Jerry M. and Gambetta, Jay M.},
  title = {Hardware-efficient variational quantum eigensolver for small molecules and quantum magnets},
  journal = {Nature},
  volume = {549},
  number = {7671},
  pages = {242--246},
  year = {2017},
  doi = {10.1038/nature23879}
}

@article{OMalley2016,
  author = {O'Malley, P. J. J. and Babbush, R. and Kivlichan, I. D. and Romero, J. and McClean, J. R. and Barends, R. and Kelly, J. and Roushan, P. and Tranter, A. and Ding, N. and Campbell, B. and Chen, Y. and Chen, Z. and Chiaro, B. and Dunsworth, A. and Fowler, A. G. and Jeffrey, E. and Lucero, E. and Megrant, A. and Mutus, J. Y. and Neeley, M. and Neill, C. and Quintana, C. and Sank, D. and Vainsencher, A. and Wenner, J. and White, T. C. and Coveney, P. V. and Love, P. J. and Neven, H. and Aspuru-Guzik, A. and Martinis, J. M.},
  title = {Scalable Quantum Simulation of Molecular Energies},
  journal = {Physical Review X},
  volume = {6},
  number = {3},
  pages = {031007},
  year = {2016},
  doi = {10.1103/PhysRevX.6.031007}
}

@article{Taube2006,
  author = {Taube, Andrew G. and Bartlett, Rodney J.},
  title = {New perspectives on unitary coupled-cluster theory},
  journal = {International Journal of Quantum Chemistry},
  volume = {106},
  number = {15},
  pages = {3393--3401},
  year = {2006},
  doi = {10.1002/qua.21198}
}

@article{Romero2018,
  author = {Romero, Jonathan and Babbush, Ryan and McClean, Jarrod R. and Hempel, Cornelius and Love, Peter J. and Aspuru-Guzik, Al{\'a}n},
  title = {Strategies for quantum computing molecular energies using the unitary coupled cluster ansatz},
  journal = {Quantum Science and Technology},
  volume = {4},
  number = {1},
  pages = {014008},
  year = {2019},
  doi = {10.1088/2058-9565/aad3e4}
}

@article{Claudino2020,
  author = {Claudino, Daniel and Wright, Jerimiah and McCaskey, Alexander J. and Humble, Travis S.},
  title = {Benchmarking Adaptive Variational Quantum Eigensolvers},
  journal = {Frontiers in Chemistry},
  volume = {8},
  pages = {606863},
  year = {2020},
  doi = {10.3389/fchem.2020.606863}
}

@article{Tang2021QubitAdapt,
  author = {Tang, Ho Lun and Shkolnikov, V. O. and Barron, George S. and Grimsley, Harper R. and Mayhall, Nicholas J. and Barnes, Edwin and Economou, Sophia E.},
  title = {{Qubit-ADAPT-VQE}: An Adaptive Algorithm for Constructing Hardware-Efficient Ans{\"a}tze on a Quantum Processor},
  journal = {PRX Quantum},
  volume = {2},
  number = {2},
  pages = {020310},
  year = {2021},
  doi = {10.1103/PRXQuantum.2.020310}
}

@article{Yordanov2021,
  author = {Yordanov, Yordan S. and Armaos, Vasilis and Barnes, Crispin H. W. and Arvidsson-Shukur, David R. M.},
  title = {Qubit-excitation-based adaptive variational quantum eigensolver},
  journal = {Communications Physics},
  volume = {4},
  number = {1},
  pages = {228},
  year = {2021},
  doi = {10.1038/s42005-021-00730-0}
}

@article{Shkolnikov2023,
  author = {Shkolnikov, Vladislav O. and Mayhall, Nicholas J. and Economou, Sophia E. and Barnes, Edwin},
  title = {Avoiding symmetry roadblocks and minimizing the measurement overhead of adaptive variational quantum eigensolvers},
  journal = {Quantum},
  volume = {7},
  pages = {1040},
  year = {2023},
  doi = {10.22331/q-2023-06-12-1040}
}

@article{Sapova2022Batched,
  author = {Sapova, Mariia D. and Fedorov, Aleksey K.},
  title = {Variational quantum eigensolver techniques for simulating carbon monoxide oxidation},
  journal = {Communications Physics},
  volume = {5},
  number = {1},
  pages = {199},
  year = {2022},
  doi = {10.1038/s42005-022-00982-4}
}

@article{Anastasiou2024Tetris,
  author = {Anastasiou, Panagiotis G. and Chen, Yanzhu and Mayhall, Nicholas J. and Barnes, Edwin and Economou, Sophia E.},
  title = {{TETRIS-ADAPT-VQE}: An adaptive algorithm that yields shallower, denser circuit ans{\"a}tze},
  journal = {Physical Review Research},
  volume = {6},
  number = {1},
  pages = {013254},
  year = {2024},
  doi = {10.1103/PhysRevResearch.6.013254}
}

@article{Ramoa2025CEO,
  author = {Ram{\^o}a, Mafalda and Anastasiou, Panagiotis G. and Santos, Luis Paulo and Mayhall, Nicholas J. and Barnes, Edwin and Economou, Sophia E.},
  title = {Reducing the resources required by {ADAPT-VQE} using coupled exchange operators and improved subroutines},
  journal = {npj Quantum Information},
  volume = {11},
  pages = {86},
  year = {2025},
  doi = {10.1038/s41534-025-01039-4}
}

@article{Sun2016QuantumEmbedding,
  author = {Sun, Qiming and Chan, Garnet Kin-Lic},
  title = {Quantum Embedding Theories},
  journal = {Accounts of Chemical Research},
  volume = {49},
  number = {12},
  pages = {2705--2712},
  year = {2016},
  doi = {10.1021/acs.accounts.6b00356}
}

@article{Rubin2016,
  author = {Rubin, Nicholas C.},
  title = {A hybrid classical/quantum approach for large-scale studies of quantum systems with density matrix embedding theory},
  journal = {arXiv preprint arXiv:1610.06910},
  year = {2016},
  eprint = {1610.06910},
  archivePrefix = {arXiv}
}

@article{Lowdin1950,
  author = {L{\"o}wdin, Per-Olov},
  title = {On the Non-Orthogonality Problem Connected with the Use of Atomic Wave Functions in the Theory of Molecules and Crystals},
  journal = {The Journal of Chemical Physics},
  volume = {18},
  number = {3},
  pages = {365--375},
  year = {1950},
  doi = {10.1063/1.1747632}
}

@article{Broyden1965,
  author = {Broyden, C. G.},
  title = {A Class of Methods for Solving Nonlinear Simultaneous Equations},
  journal = {Mathematics of Computation},
  volume = {19},
  number = {92},
  pages = {577--593},
  year = {1965},
  doi = {10.1090/S0025-5718-1965-0198670-6}
}

\end{document}